\documentclass[12pt]{article}
\usepackage{amsmath}
\usepackage{amsfonts}
\usepackage{amssymb}
\usepackage{latexsym}
\usepackage{graphicx}
\usepackage{caption}
\usepackage{subcaption}

\gdef\ffrac#1#2{\textstyle\frac{#1}{#2}\displaystyle}
\gdef\be{\begin{equation}}
\gdef\ee{\end{equation}}

\begin{document}
\setcounter{page}{0} \topmargin 0pt
\renewcommand{\thefootnote}{\arabic{footnote}}
\newpage
\setcounter{page}{0}

\begin{titlepage}

\begin{center}
{\Large {\bf Quantum Revivals in Conformal Field Theories in Higher Dimensions}}\\

\vspace{2cm}
{\large John Cardy$^{a,b}$\\}  \vspace{0.5cm} {\em $^{a}$Department of Physics\\
University of California, Berkeley CA 94720, USA\\}  \vspace{0.2cm} {\em $^{b}$All Souls College, Oxford OX1 4AL, UK\\}

\vspace{2cm}

\end{center}

\vspace{1cm}

\begin{abstract}
\noindent We investigate the behavior of the return amplitude ${\cal F}(t)= |\langle\Psi(0)|\Psi(t)\rangle|$  following a quantum quench in a conformal field theory (CFT) on a compact spatial manifold of dimension $d-1$ and linear size $O(L)$, from a state $|\Psi(0)\rangle$ of extensive energy with short-range correlations. After an initial gaussian decay ${\cal F}(t)$ reaches a plateau value related to the density of available states at the initial energy. However for $d=3,4$ this value is attained from below after a single oscillation. For a holographic CFT the plateau persists up to times at least $O(\sigma^{1/(d-1)} L)$, where $\sigma\gg1$ is the dimensionless Stefan-Boltzmann constant. On the other hand for a free field theory on manifolds with high symmetry there are typically revivals at times $t\sim\mbox{integer}\times L$. In particular, on a sphere $S_{d-1}$ of circumference $2\pi L$, there is an action of the modular group on ${\cal F}(t)$ implying structure near all rational values of $t/L$, similarly to what happens for rational CFTs in $d=2$.

\end{abstract}

\end{titlepage}

\section{Introduction}\label{sec1}

Quantum revivals, when after some finite time $t$ the initial state is wholly or partially recovered, are generally a property of simple systems with only a few degrees of freedom. The simplest example is the harmonic oscillator, where, since the energy gaps are quantized in integer multiples of $\hbar\omega$, the return amplitude
$$
{\cal F}(t)=|\langle\Psi(0)|\Psi(t)\rangle|=|\langle\Psi(0)|e^{-iHt}\Psi(0)\rangle|=|\sum_n|\langle n|\Psi(0)\rangle|^2e^{-in\omega t}|
$$
is exactly unity at times $t=(2\pi/\omega)\times$ integer. More generally, such full or partial revivals at which 
${\cal F}=O(1)$ should occur only if the ratios of energy gaps happen to be rational numbers with a finite common denominator, or if the initial state happens to couple to only such energy eigenstates. In more complex quantum systems such coincidences are rare, and one expects to see no revivals at which ${\cal F}(t)$ is $O(1)$. Indeed, the overlap should in general vanish roughly as $N^{-1/2}$ where $N$ is the dimension of the Hilbert space. For a system with an extensively large number of degrees of freedom, this implies an exponential vanishing of ${\cal F}(t)$ with the volume.

\begin{figure}[ht]
\centering
\begin{subfigure}{0.5\textwidth}
\centering
\includegraphics[width=0.9\linewidth]{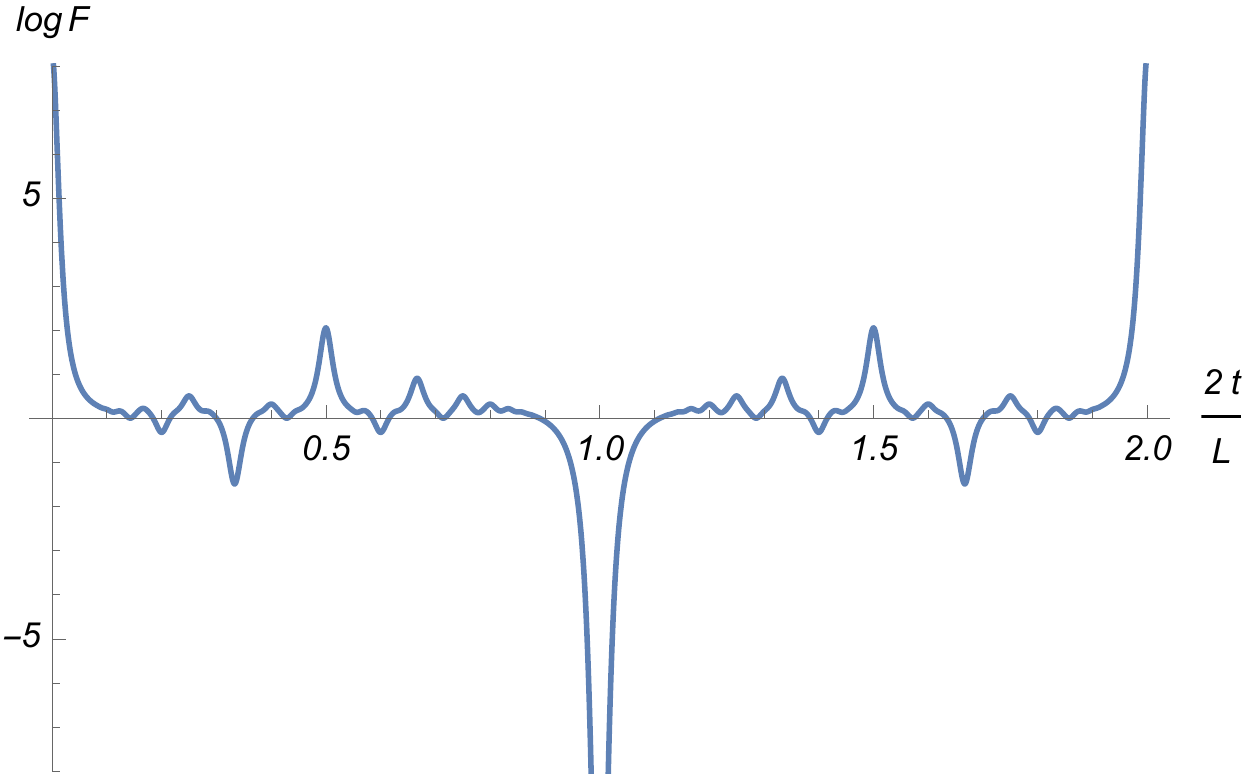}
\caption{\small $2\pi\beta/L=0.1$}
\end{subfigure}%
\begin{subfigure}{0.5\textwidth}
\centering
\includegraphics[width=0.9\linewidth]{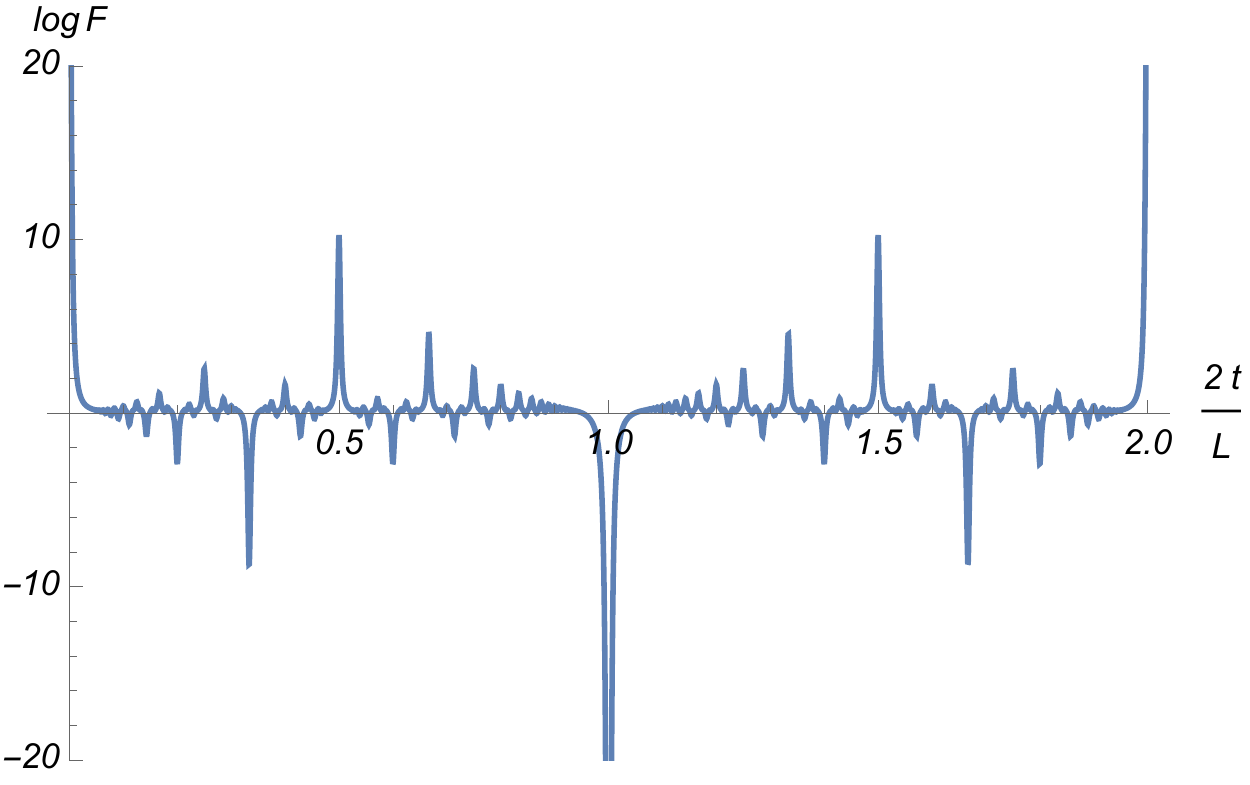}
\caption{\small $2\pi\beta/L=.02$}
\end{subfigure}
\caption{\small Logarithmic return amplitude for the 2d Ising field theory quenched from the disordered phase. There are full revivals at even integer $2t/L$. At odd values there is in fact destructive interference between even and odd values of the conformal dimensions. Partial revivals at simple rational values are also visible, more pronounced ar larger values of $L/\beta$. }
\label{Fig2d}
\end{figure}

Even for such extensive systems, however, there are exceptions for which quantum revivals may occur. This can happen if there is a spectrum-generating algebra with ladder operators analogous to those in the harmonic oscillator, for example the Virasoro (or extended) algebras in rational CFTs on a circle for $d=2$, or for massless free scalar field theory on the torus or sphere in higher dimensions. In a previous paper \cite{C1} the $d=2$ case was investigated for the case when the initial state has short-range correlations and area-law entanglement. This scenario is often called a quantum quench, because the initial state approximates the ground state of a pre-quench gapped hamiltonian, while the dynamics proceeds according to the gapless hamiltonian of a CFT. It was found there that, for a rational CFT, quantum revivals at which ${\cal F}=O(1)$  generally occur at times such that $2t/L=$ integer, where $L$ is the circumference of the circle, although full revivals occur only at certain multiples of this.  Moreover it was shown that, for a particular initial state of the form 
\begin{equation}\label{ccstate}
|\Psi(0)\rangle=e^{-(\beta/4)H_{CFT}}|B\rangle\,,
\end{equation}
where $|B\rangle$ is a conformal boundary state, ${\cal F}$ may be expressed in terms of the partition function on an annulus, continued to complex values of the modulus. This may be written as a linear combination of Virasoro characters which are functions of $q=e^{2\pi i\tau}=e^{-2\pi(\beta+2it)/L}$, where $\beta$ is the effective inverse temperature of the initial state. The fact that these transform simply under the modular group generated by $\tau\to\tau+1$ and $\tau\to-1/\tau$ implies that there are not only revivals at integer values of $2t/L$, but echoes of these near all rational values of this ratio. An example is shown in 
Fig.~\ref{Fig2d}.

These revival echoes are superimposed on a plateau value ${\cal F}(t)\sim\exp(-\pi cL/12\beta)$. In 2d this is attained from above according to the formula \cite{C1}
\be\label{2ddecay}
{\cal F}(t)\sim\exp\big(-(\pi c/3)Lt^2/\beta(\beta^2+4t^2)\big)\,.
\ee
 This follows from a well-controlled analytic continuation of the Casimir contribution to the free energy of a strip of width $\beta/2+it$ and length $L\gg|\beta+2it|$. As we later discuss in detail for general $d$, it may also be seen as a consequence of interference between high-energy intermediate states of the CFT and the well-known form \cite{C2} for their density (as well as their matrix elements.) Eq.~(\ref{2ddecay}) is generally valid only for $t\ll(L\beta)^{1/2}$, although for a rational CFT its form is echoed at all the revivals and partial revivals. However, as we shall also point out for general $d$, for a holographic CFT, which we loosely take to mean $c\gg1$ with a sparse density of operators with conformal dimensions which are $o(c)$, this asymptotic form (and thus its plateau limit) persists at least as far as $t=O(cL)$, that is, there are no quantum revivals with $t/L=O(1)$. This is of course consistent with the AdS/CFT interpretation that in the quench scenario an in-falling initial energy/mass density will result in the formation of a black hole and purely thermal behavior (although the detailed dynamics of this process may be more complicated). \cite{BH}
 
In this paper we generalize these results to CFTs in space-time dimensions $d>2$. In Sec.~\ref{sec2}  we show, for a general CFT on a compact manifold $M$ that that the initial 
gaussian decay of ${\cal F}(t)$ is always present, but that the approach to the plateau value is via oscillations, the number of which increases with increasing $d$. In Sec.~\ref{sec3} we consider the simplest CFT of all, a massless scalar field theory, in the case where $M$ has the most symmetry, \em i.e. \em is a sphere $S_{d-1}$ of circumference $L$ (i.e. radius $L/2\pi$ when embedded in ${\mathbb R}^d$). Once again we assume an initial state of the form (\ref{ccstate}) with $|B\rangle$ now corresponding to a Dirichlet boundary condition. The return amplitude is then once again related to a partition function, on $S_{d-1}\times[0,\beta/2]$, continued to complex $\beta\to\beta+2it$. We further argue that this partition function is simply related to that on $S_{d-1}\times S_1$, which, for a massless free scalar CFT was considered in detail in Ref.~\cite{JC91}. It may be computed either directly, by computing the grand partition function from the single-particle spectrum, or by a conformal mapping which takes the imaginary time
$\tau$ to a radial coordinate $r=e^\tau$ in $R^d$, whence, by the operator-state equivalence in CFT,
$$
Z(S_{d-1}\times S_1)\propto {\rm Tr}\,e^{-\beta H_{CFT}}=\sum_\Delta q^\Delta\,,
$$
where $q=e^{-2\pi\beta/L}$ and the sum is over the spectrum of scaling dimensions $\Delta$ of the CFT. For a free field theory this is straightforward to compute by enumerating all operators all of which have canonical integer (or half-integer) dimensions. This last fact immediately implies the existence of complete quantum revivals at integer values of $2t/L$ (or $t/L$).

However, in \cite{JC91} it was pointed out that, for $d$ even, this partition function has simple properties under modular transformations which interchange $\beta\leftrightarrow L$ (For $d$ odd a more complicated version holds for antiperiodic boundary conditions around the $S_1$ cycle). While for a 2d CFT this follows from interchanging two cycles of a torus and should hold for any CFT, for $d>2$ there is no such general principle and it appears to be a particular property of free field theories. Nevertheless, we may use it, as in \cite{C1}, to infer peak structure in $\log{\cal F}(t)$ near all rational values of $2t/L$, not only integers. Near each such value there will be an image, or echo, of the initial oscillating decay. On the other hand we find that for $d$ odd there is destructive interference at
half-integer $2t/L$, and echoes of these negative peaks then also appear at other rational values. As we shall see, these negative peaks are in fact related to the behavior of a fermionic partition function.

\section{Initial state and initial decay}\label{sec2}
We  consider a CFT on a $(d\!-\!1)$-dimensional compact spatial manifold $M$ of characteristic size $L$, evolving from an initial state $|\Psi(0)\rangle$ at time $t=0$. This state is taken to have short-range correlations such as would occur in the non-degenerate ground state of a massive quantum field theory. Another way of stating this is to require that the mutual information between two widely separated regions should decay exponentially over distances smaller than $L$.
 
 In 2d,  in \cite{CC} a choice was made for such a state of the form given in Eq.~\ref{ccstate}, where $|B\rangle$ is a conformally invariant boundary state. This was motivated partly by renormalization group considerations that all boundary states should be equivalent to a conformal state on large distance scales, but more practically for reasons of analytic tractability.  However, as was argued in \cite{CC}, and shown more completely in \cite{JCGGE}, in 2d the state (\ref{ccstate}) has the property that it \em thermalizes\em, that is, although the whole system remains in a pure state, the reduced density matrix of any subsystem becomes exponentially close to the reduced density matrix of a thermal Gibbs ensemble $\propto e^{-\beta H_{CFT}}$ after a finite time (essentially the time it takes for a light signal to cross the subsystem). Since generic systems are usually assumed to thermalize, this is therefore a useful idealized model for how this happens in a solvable model. 
 
 However, another way to view the state in (\ref{ccstate}) is to recognize that $H_{CFT}=\int T_{00}dx$ is the space integral of one of the leading irrelevant operators which may act on $|B\rangle$. This suggests that a more general initial state may be constructed by adding to the exponent in (\ref{ccstate}) all possible other irrelevant operators.  For a 2d CFT this was argued in \cite{JCGGE} to lead to a generalized Gibbs ensemble (GGE), which takes into account all the local conservation laws of the CFT (essentially integrals of local holomorphic and anti-holomorphic currents.) In general it is has been argued that in such integrable models the stationary state should be described
 by a GGE \cite{GGE1}, and exactly which conservation laws should be included has been the subject of considerable investigation \cite{GGE2}.
 
 However, for $d>2$ we expect most CFTs to be non-integrable and therefore to thermalize to a Gibbs ensemble after a quench from any reasonable state. For this reason we may consider the analog of (\ref{ccstate}) as an initial state, where $|B\rangle$ is now a boundary state on $M$, defined by $T_{0j}|B\rangle=0$, where $j$ is a coordinate labeling $M$. We assume throughout that the volume of $M$ is $O(L^{d-1})$ with $L\gg\beta$. 
 We argue in the Appendix that, for a free field theory with $|B\rangle$ corresponding to Dirichlet boundary conditions, this state should thermalize in the above sense. For holographic CFTs with large central charge one also expects this as a consequence of the Hawking radiation after 
formation of a black hole \cite{BH} in AdS$_{d+1}$. However for 2d CFTs and free field theory in $d>2$ the thermalization argument technically relies on a use of the method of images and analyticity arguments which do not obviously apply to a general $d>2$ CFT.

 The (conserved) energy of the state (\ref{ccstate}) is 
 $$
\langle\Psi| H|\Psi\rangle=\frac{\langle B|e^{-(\beta/4)H}He^{-(\beta/4)H}|B\rangle}
 {\langle B|e^{-(\beta/4)H}e^{-(\beta/4)H}|B\rangle}
 =-2(\partial/\partial\beta)\log Z(M\times[0,\beta/2])\,,
 $$
 where $Z(M\times[0,\beta/2])$ is the CFT partition function on $M$ times the interval $[0,\beta/2]$ with boundary conditions $B$, and from now on we drop the suffix on $H_{CFT}$. For $L\gg\beta$ we expect that
\be\label{Casimir}
 \log Z(M\times[0,\beta/2])\sim\tilde\sigma\,\frac{{\rm Vol}(M)}{(\beta/2)^{d-1}}\,,
 \ee
 which (for $d=3$) is the Casimir free energy between two parallel plates of separation $\beta/2$. Here $\tilde\sigma$ is a universal constant depending on the CFT and the boundary condition $B$. Note that for a curved manifold $M$ we expect also curvature dependent terms in $\log Z$. These should be proportional to the integrated curvature and so down by a relative factor of $(\beta/L)^2$ compared with the leading term.  
 
 For a 2d CFT, and for a free theory for $d>2$ with Dirichlet boundary conditions, $\tilde\sigma=\frac12\sigma$ where 
 $$
\log Z(M\times S_1(\beta))\sim\sigma\,\frac{{\rm Vol}(M)}{\beta^{d-1}}
$$
is (minus) the free energy at finite temperature $\beta^{-1}$. $\sigma$ is proportional to the Stefan-Boltzmann constant of the CFT and determines the asymptotic  behavior of the density of states. For these cases 
$\langle\Psi| H|\Psi\rangle$ is equal to the mean energy $\propto\sigma T^d\,{\rm Vol}(M)$ of the CFT at temperature $T=\beta^{-1}$. 

This relation between $\tilde\sigma$ and $\sigma$ is a consequence of the method of images being applicable to correlators of the stress tensor, for all 2d CFTs and for free theories in $d>2$. However there seems to be no reason for it to hold more generally for $d>2$ (although they should still be proportional), and in this case we take $\beta$ to be simply a length scale parametrizing the initial state $|\Psi(0)\rangle$. 

Let us now consider the return amplitude 
\be\label{ret}
{\cal F}(t)=|\langle B|e^{-(\beta/4)H}e^{-itH}e^{-(\beta/4)H}|B\rangle\,.
\ee
At this point the utility of the choice (\ref{ccstate}) becomes apparent, since this is simply the ratio of partition functions
\begin{equation}\label{Fratio}
{\cal F}(t)=\left|\frac{Z(M\times[0,\beta/2+it])}{Z(M\times[0,\beta/2])}\right|
\end{equation}
Note that the analytic continuation is always possible, since if we insert a complete set of energy eigenstates
\be\label{rhoE}
Z(M\times[0,\beta/2+it])=\int \rho(E)|\langle B|E\rangle|^2e^{-(\beta/2+it)E}dE\,,
\ee
where $\rho(E)$ is the density of states, and this should converge for all $\beta>0$. (Note that in principle there are also UV-divergent terms in $\log Z$ proportional to the $d$-dimensional space-time volume and also the $d$-dimensional volume of $M$. However when suitably regularized these cancel in (\ref{Fratio}).) 

The denominator in (\ref{Fratio}) is dominated for $L\gg\beta$ by the Casimir term (\ref{Casimir}). This should continue to be the case for sufficiently small $t$. In that case
$$
\log{\cal F}(t)\sim \tilde\sigma{\rm Vol}(M)\,{\rm Re}\,\left[\frac1{(\beta/2+it)^{d-1}}-\frac1{(\beta/2)^{d-1}}\right]\,.
$$
For $d=2$ this gives the result (\ref{2ddecay}). For $d>2$, however, the behavior is more complicated.
Indeed  we have
\begin{eqnarray*}
\log{\cal F}(t)-\log{\cal F}(\infty)&\propto&\frac{\beta^2-4t^2}{(\beta^2+4t^2)^2}\quad(d=3)\,,\\
&\propto&\frac{\beta(\beta^2-12t^2)}{(\beta^2+4t^2)^3}\quad(d=4)\,.
\end{eqnarray*}
Thus in these two cases the height above the plateau value changes sign at values of $t=O(\beta)$ and, unlike the case $d=2$, the asymptotic value is reached from below. These are clearly visible in the examples in Figs.~(\ref{Fig4da}, \ref{Fig3d}). For larger values of $d$ there are more oscillations, and
it may be shown that in general the number of sign changes is $[(d-1)/2]$.

We now discuss the domain of applicability of these results. This may be ascertained by estimating the corrections to the Casimir term in the free energy. If $M$ has sufficient symmetry (for example, itself contains an $S_1$ factor) we may quantize along this direction so that `space' has dimensions $O(L^{d-2}\times\beta)$ and imaginary `time' dimension $O(L)$. In this picture, the ground state energy is extensive in the $(d-2)$ directions, so the free energy should be $\propto L^{d-2}\cdot L$, and the universal finite-size term therefore $\propto1/\beta^{d-1}$. This accounts for the leading Casimir term in (\ref{Casimir}). The corrections to this, however, should come in this picture from excited states which should have a finite-size gap dependence $\propto \beta^{-1}$. Thus the corrections to the partition function should have the form
$$
Z(M\times[0,\beta/2])=e^{\tilde\sigma{\rm Vol}(M)/(\beta/2)^{d-1}}\left(1+O\big(e^{-{\rm const.}L/\beta}\big)\right)\,,
$$
just as for $d=2$ \cite{C1}. In the next section we find this to be the case for a free field theory on $S_{d-1}$, as a consequence of modular symmetry, the only difference being a power-law pre-factor. This may be understood in terms of the excited states forming a continuum as a result of the additional large $(d-2)$ dimensions.

Assuming this to be the case in general, we can now assert that for $t>0$ the Casimir term dominates as long as
$|e^{-L/(\beta+2it)}|\ll1$, which translates into $t\lesssim(\beta L)^{1/2}$, just as in 2d \cite{C1}. Note that this time scale is greater than the width $O(\beta)$ of the initial gaussian and any transient oscillations. Thus the plateau behavior should always be observable as long as $L\gg\beta$.

However the plateau behavior may persist longer than this in holographic theories. The Casimir term in (\ref{Casimir}) implies for the high-energy behavior of the density of states (times the mean square matrix element) in (\ref{rhoE}) 
$$
\rho(E)\overline{|\langle B|E\rangle|^2}\sim\int_C\frac{d\beta}{2\pi i}e^{\beta E}e^{\tilde\sigma V/\beta^{d-1}}\,,
$$
where $V={\rm Vol}(M)$ and the $\beta$ contour $C$ is parallel to the imaginary axis. A steepest descent calculation then gives, apart from prefactors,
\be\label{asymp}
\rho(E)\overline{|\langle B|E\rangle|^2}\sim e^{(d/(d-1))((d-1)\tilde\sigma V)^{1/d}E^{1-1/d}}\,.
\ee
 If instead we apply the same method to $Z(M\times S_1)$, with $\tilde\sigma\to\sigma$, this gives the $d$-dimensional generalization of the well-known formula for the asymptotic density of states in a CFT. In 2d \cite{BCN,Affleck}, $\sigma=\pi c/6$ and\footnote{Note that the annulus has width $\beta/2$ so the denominator is 12 rather than the usual 24 \cite{BCN}.} $\tilde\sigma=\pi c/12$. Note that $\rho(E)$ in the above is not the full density of states, but rather that of those which have significant overlap with $|B\rangle$, which accounts for the reduction $\tilde\sigma<\sigma$. 
 
 On a compact manifold $M$, the energies $E$ in the above should always in fact be discrete, and quantized in units of $L^{-1}$. If $M=S_{d-1}$, as will be considered in the next section, they have the form $E=E_0+2\pi\Delta/L$, where the $\Delta$s are the scaling dimensions of the (scalar) operators of the CFT, and $E_0\sim-2\pi\bar\sigma/L$, where $\bar\sigma>0$ measures the ground state energy in the limit when $L\ll\beta$.  In that case, (\ref{asymp}) is generally valid as long as
 $\Delta\gg\bar\sigma$. For \em holographic \em CFTs, however, all these Casimir coefficients $(\sigma,\tilde\sigma, \bar\sigma)$ are $\gg1$ and $\rho(E)$ is vanishingly small on this scale for $\Delta<\bar\sigma$. In that case, (\ref{asymp}) holds more generally, for $\Delta\geq\bar\sigma\gg1$. Strictly speaking, this has been shown only for large $c$ 2d CFTs \cite{Hart}, based on modular invariance, but this property should also extend to $d>2$ on the basis of holographic arguments. 
 
 Assuming this wider applicability of (\ref{asymp}), we may substitute it back into (\ref{rhoE}) and once again perform a steepest descent calculation. In that case, for $t=0$ we of course recover the form (\ref{Casimir}), but now the approximation is valid so long as $\tilde\sigma(L/\beta)^{d-1}\gg1$. For non-zero $t\gg\beta$, this translates into the result that the plateau persists up to times
 $$
 t\lesssim{\tilde\sigma}^{1/(d-1)}L\,.
 $$
 For such CFTs with $\tilde\sigma\gg1$, this rules out any revivals occurring at times $t=O(L)$. This is of course consistent with the post-quench dynamics in such theories being holographically equivalent to the formation of a black hole \cite{BH}.
 
 Finally in this section we note that the plateau value of the overlap
 $$
|\langle\Psi(0)|\Psi(t)\rangle|\sim e^{-\tilde\sigma{\rm Vol}(M)/\beta^{d-1}}
$$
 should be compared to the density of available states, given at the saddle-point by
 $$
 \rho^{\rm av}(E)=\rho(E)\overline{|\langle B|E\rangle|^2}\sim e^{d\tilde\sigma{\rm Vol}(M)/\beta^{d-1}}\,.
 $$
 Note the additional factor of $d$ in the exponent, coming from the saddle-point computation. If both 
 $|\Psi(0)\rangle$ and $|\Psi(t)\rangle$ are taken to be random states in an energy range around the saddle-point value, we would expect
 $$
 |\langle\Psi(0)|\Psi(t)\rangle|\sim\rho^{\rm av}(E)^{-1/2}\,.
 $$
 While this works in 2d,  for $d>2$ the overlap is much larger than expected on the basis of this simple argument.

 \section{Revivals in free field theory}\label{sec3}
 
 To go beyond these general results for $d>2$ we must specify a particular CFT and a particular manifold $M$. 
 In this section we consider the case when $M=S_{d-1}$, with equatorial circumference $L$,  and the CFT is that of a massless free scalar, given by the euclidean action
 
\be\label{ff}
 S=\ffrac12\int_{S_{d-1}}\int\big((\partial^\mu\phi\partial_\mu\phi)+\xi R\,\phi^2\big)\sqrt g\,d^{d-1}\!x\,d\tau\,,
 \ee
 where $\xi$ parametrizes the coupling to the local scalar curvature $R$. In order for this to be conformally invariant with $\phi$ transforming according to its canonical dimension $d/2-1$, $\xi$ should be set to $(d-2)/4(d-1)$.
 For a sphere this means that $\xi R=\lambda(2\pi/L)^2$ with $\lambda=(d/2-1)^2$.
 It was pointed out in \cite{JC91} that it is only for this value that $Z(S_{d-1}\times S_1)$ enjoys simple modular properties. We shall see later that this is also the case for quantum revivals. 
 
 We consider a quench in this theory from the initial state $|\Psi(0)\rangle\propto e^{-(\beta/4)H}|D\rangle$, where $D$ denotes the Dirichlet boundary state. Although a free field theory on $S_{d-1}$ has the property that all the mode numbers are conserved, so we would expect the long time behavior to be given by a GGE \cite{freefieldGGE}, as we show in the Appendix 
 this particular state has the property that subsystems thermalize to those of a simple Gibbs ensemble. 
 
Having justified the use of this initial state, we turn to the return amplitude, which has the form (\ref{Fratio}). 
Although we need to compute the partition function on $M\times[0,\beta/2]$, we begin by summarizing the results of Ref.~\cite{JC91} for the finite-temperature grand partition function $Z(M\times S_1(\beta))$ since they will turn out to be simply related.

The hamiltonian generating translations around the $S_1$ is that for an assembly of bosons
$$
H=\sum_{l=0}^\infty\sum_m\big(\sqrt{l(l+d-2)+\lambda}\,a^{\dag}_{l m}a_{lm}\big)\,,
$$ 
where $l$ labels representations of O$(d)$ with Casimir $l(l+d-2)$ and $m$ the states in each representation. Denoting the degeneracy of each representation by $D_d(l)$, the partition function is 
\be\label{ZNC}
Z_d=\prod_{l=0}^\infty\frac1{\big(1-e^{-(2\pi\beta/L)\sqrt{l(l+d-2)+\lambda}}\big)^{D_d(l)}}\,.
\ee
Note that in the conformally coupled case this simplifies to
\be\label{ZCC}
Z^c_d=\prod_{l=0}^\infty\frac1{\big(1-q^{l+d/2-1}\big)^{D_d(l)}}\,,
\ee
where we have introduced the modulus $q=e^{-2\pi\beta/L}$ as in 2d. 
Examples are
$$
Z^c_2=\prod_{l=1}^\infty\frac1{\big(1-q^{l}\big)^{2}}\,,\quad
Z^c_3=\prod_{l=0}^\infty\frac1{\big(1-q^{l+1/2}\big)^{2l+1}}\,,\quad
Z^c_4=\prod_{l=0}^\infty\frac1{\big(1-q^{l+1}\big)^{(l+1)^2}}\,.
$$
Note that in 2d the $l=0$ zero mode must be subtracted. The expressions above are normalized so that the contribution of the vacuum state gives 1. In Ref.~\cite{JC91} it was shown how this can be reinstated by studying the modular properties of $Z_d$, which will also be discussed below.

Also note that $Z_2^{1/2}$ is the well-known generating function for partitions of integers, and $Z_3$ is related to that of plane partitions, whose generating function is\cite{planepart}
$$
Z_{\text{plane partitions}}=\prod_{l=0}^\infty\frac1{\big(1-q^{l+1}\big)^{l+1}}\,.
$$

In Ref.~\cite{JC91} $Z_d^c$ was related to a similar enumeration problem. 
The metric on $S_{d-1}\times{\mathbb R}$ is
$$
ds^2=+d\tau^2+(L/2\pi)^2d\Omega\,.
$$ 
Setting $\tau=(L/2\pi)\log r$ this becomes
$$
ds^2=(L/2\pi r)^2\big(dr^2+r^2d\Omega\big)\,,
$$
so is conformally equivalent to the euclidean metric on ${\mathbb R}^d$. The generator of translations in $\tau$ is proportional to the generator of scale transformations of $r$. Of course this is simply the $d$-dimensional version of radial quantization. It shows in particular that, up to possible anomaly terms, to be discussed later,
\be\label{op}
Z^c\big(S_{d-1}\times S_1(\beta)\big) \propto Z^c_d(q)=\sum_\Delta q^\Delta\,,
\ee
where the sum is over all the scaling dimensions of the CFT, and $q=e^{-2\pi\beta/L}$. For the free field theory, this will be true only for the conformally coupled case. 

For the free scalar field theory in $d$ dimensions, a list of independent operators is
\be\label{list}
\big(\partial_1^{n_1^{(1)}}\partial_2^{n_2^{(1)}}\ldots\partial_d^{n_d^{(1)}}\!\!\phi\big)
\big(\partial_1^{n_1^{(2)}}\partial_2^{n_2^{(2)}}\ldots\partial_d^{n_d^{(2)}}\!\!\phi\big)\ldots\,,
\ee
modulo the equation of motion $\sum_{i=1}^d\partial_i^2\phi=0$. This allows us to express all the terms with $n_1^{(j)}\geq2$ as linear combinations of other operators,s o that we may restrict to $n_1^{(j)}=0,1$. The operator in (\ref{op}) has scaling dimension $\Delta=\sum_j(d/2-1+n_1^{(j)}+\cdots+n_d^{(j)})$. The generating function is then
$$
Z_d(q)=\prod_{n_1=0}^1\prod_{n_2=0}^\infty\cdots\prod_{n_d=0}^\infty
\frac1{1-q^{d/2-1+n_1+n_2+\cdots+n_d}}\,.
$$
Defining 
$$
f_d(l)=\sum_{n_i}\delta\left(\sum_{i=2}^dn_i-l\right)=\frac{(d-l-2)!}{l!(d-2)!}\,,
$$
we then have
\begin{eqnarray}
Z_d(q)&=&\prod_{l=0}^\infty\frac1{(1-q^{d/2-1+l})^{f_d(l)}}\,\prod_{l=0}^\infty\frac1{(1-q^{d/2-1+l+1})^{f_d(l)}}\\
&=&
\prod_{l=0}^\infty\frac1{(1-q^{d/2-1+l})^{D_d(l)}}\,,\label{Zd}
\end{eqnarray}
where 
$$
D_d(l)=f_d(l)+f_d(l-1)=\frac{(d+2l-2)(d+l-3)!}{l!(d-2)!}
$$
is in fact just the dimension of the representation of O$(d)$ with `total angular momentum' $l$, that is Casimir
$l(l+d-2)$: $D_3(l)=2l+1$, $D_4(l)=(l+1)^2$, etc. This may be checked by expanding each factor in (\ref{Zd}) to lowest order, corresponding to including only operators with a single field $\phi$. This counts all possible multinomials of degree $l$ in $(\partial_1,\ldots,\partial_d)$ modulo the equation of motion, which give a basis for an irreducible representation of O$(d)$.

Of course it is not surprising to find the operator content of the $d$-dimensional CFT labeled by representations of 
O$(d)$. What is special for the conformally coupled free scalar is that the degeneracies are much greater than than that expected on the basis of this symmetry alone. 

We now return to the case of interest 
$$
Z(S_{d-1}\times[0,\beta/2])=\langle D|e^{-(\beta/2)H}|D\rangle\,.
$$
In principle this could be evaluated by inserting a complete set of eigenstates of $H$, one corresponding to each operator in (\ref{list}). However we would then be faced with computing the overlaps between these states and $|D\rangle$. Moreover only states with total angular momentum $l=0$ will contribute, since $|D\rangle$ is rotationally invariant. 

Fortunately this task may be avoided by the following observation. Going back to $S_{d-1}\times S_1$, we may also compute the partition function by decomposing the field into normal modes in the usual way. Then
\begin{eqnarray}
\log Z(S_{d-1}\times S_1)&=&-\ffrac12{\rm Tr}\log(-\nabla^2+\xi R)\nonumber\\
&\sim&-\ffrac12\sum_{l=0}^\infty\sum_{n=-\infty}^\infty\log\big((l(l+d-2)+\lambda)(2\pi/L)^2+n^2(2\pi/\beta)^2\big)\,.
\label{Zlog}
\end{eqnarray}
The sum is of course divergent and must be regularized, for example using zeta-function methods. As discussed in \cite{JC91} the regularization gives rise to the ground state energy term in the limit $\beta/L\gg1$.

In the above the $\tau$-dependence of the modes is $e^{2\pi in\tau/\beta}$, or, in terms of standing waves, 
$\big((\cos(2\pi n\tau/\beta),\sin(2\pi n\tau/\beta)\big)$ with $n\geq0$. The cosine modes satisfy Neuman boundary conditions at $\tau=0$. However the sine modes satisfy Dirichlet conditions at $\tau=0$ and $\beta/2$, as needed. Therefore (\ref{Zlog}) also gives the logarithmic partition function on $S_{d-1}\times[0,\beta/2]$ if we divide by a factor of 2. (This argument ignores the $n=0$ mode but this does not depend on $\beta$ and so the difference disappears in the ratio (\ref{Fratio}).)
We conclude that
$$
Z(S_{d-1}\times[0,\beta/2])\propto Z_d(q)^{1/2}\,,
$$
with $Z_d$ given by (\ref{Zd}) and again $q=e^{-2\pi\beta/L}$. 

This is of course implies the result $\tilde\sigma=\frac12\sigma$ for the coefficient of the Casimir energy for $L/\beta\gg1$. This fact is well known for 2d CFTs but it also holds for free scalar fields for $d>2$. 

\begin{figure}[ht]
\centering
\begin{subfigure}{0.5\textwidth}
\centering
\includegraphics[width=0.9\linewidth]{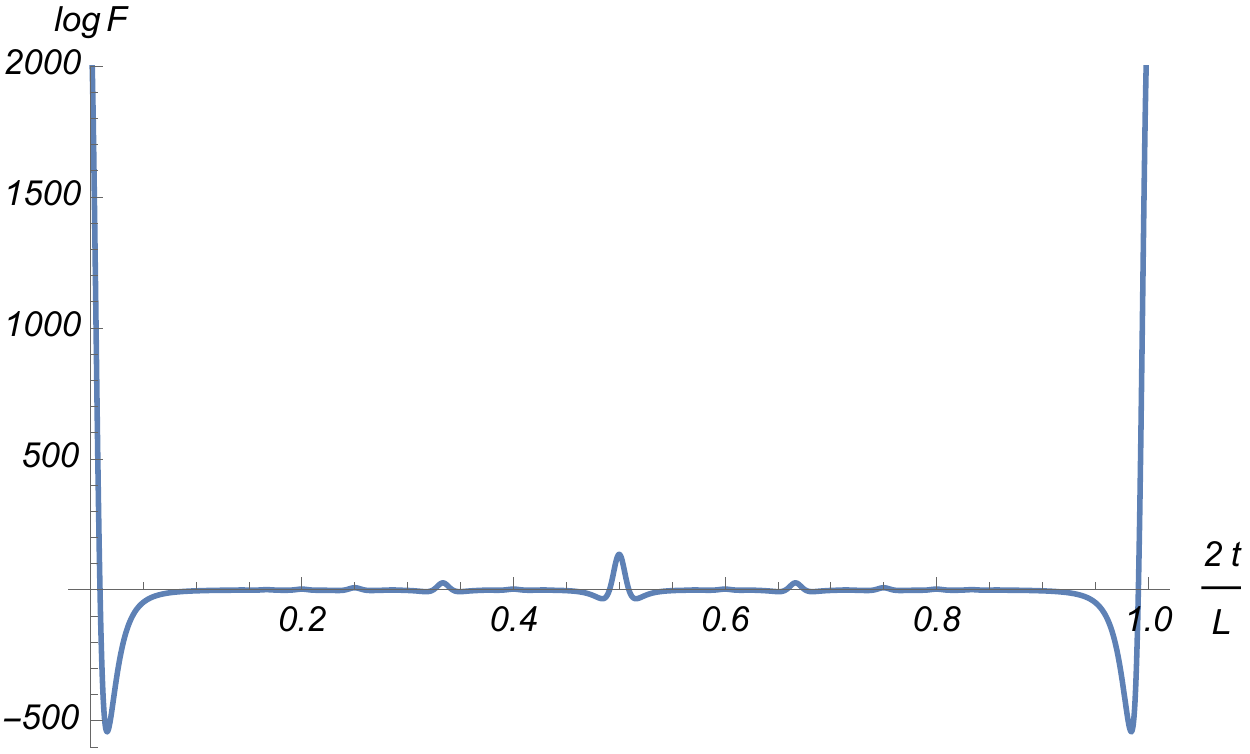}
%\caption{$2\pi\beta/L=0.1$}
\end{subfigure}%
\begin{subfigure}{0.5\textwidth}
\centering
\includegraphics[width=0.9\linewidth]{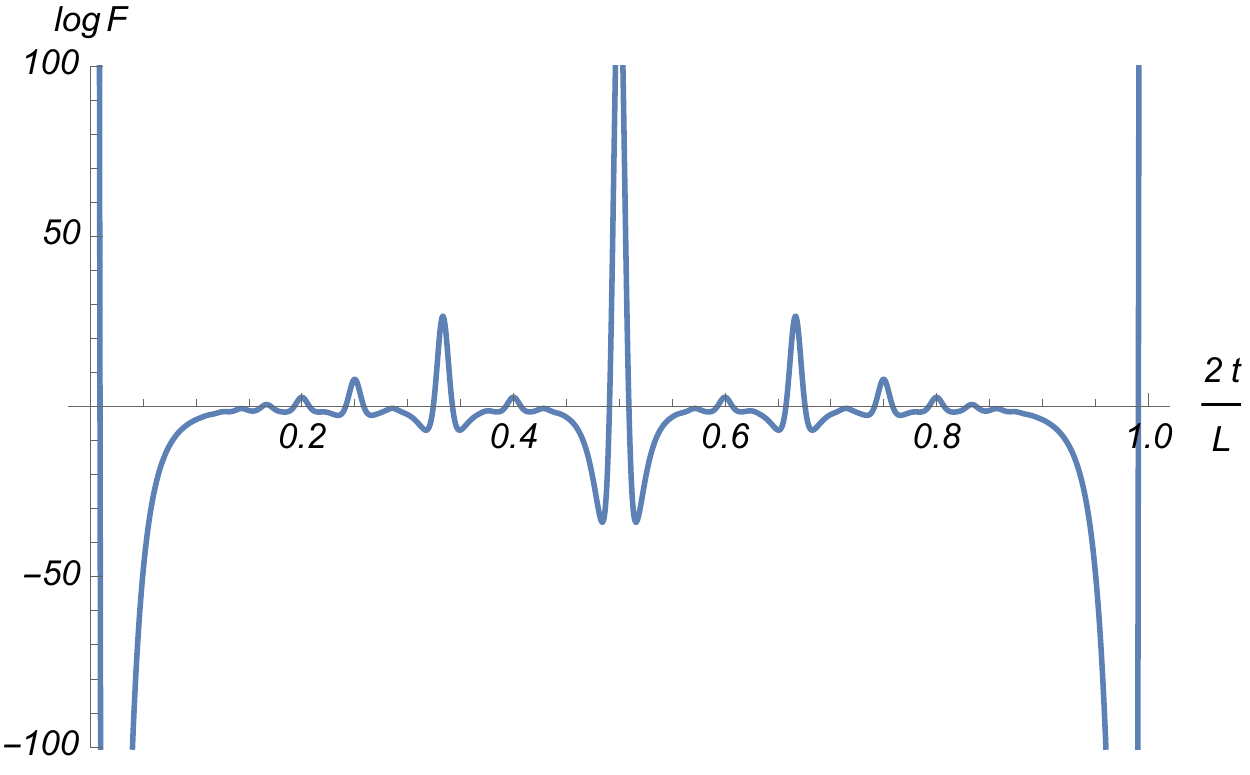}
%\caption{$2\pi\beta/L=.02$}
\end{subfigure}
\caption{\small Logarithmic return amplitude for a conformally coupled free scalar field on $S_3$, with $2\pi\beta/L=0.1$.
The initial gaussian decay is almost invisible, but the single oscillation before the approach to the plateau value is clear. Note that in this and all subsequent figures the vertical scale has been shifted, by evaluating only the numerator in (\ref{Fratio}), so that the plateau is at zero height.  In the right-hand figure the vertical axis has been expanded to show the partial revivals.}\label{Fig4da}
\end{figure}

\begin{figure}[ht]
\centering
\begin{subfigure}{0.5\textwidth}
\centering
\includegraphics[width=0.9\linewidth]{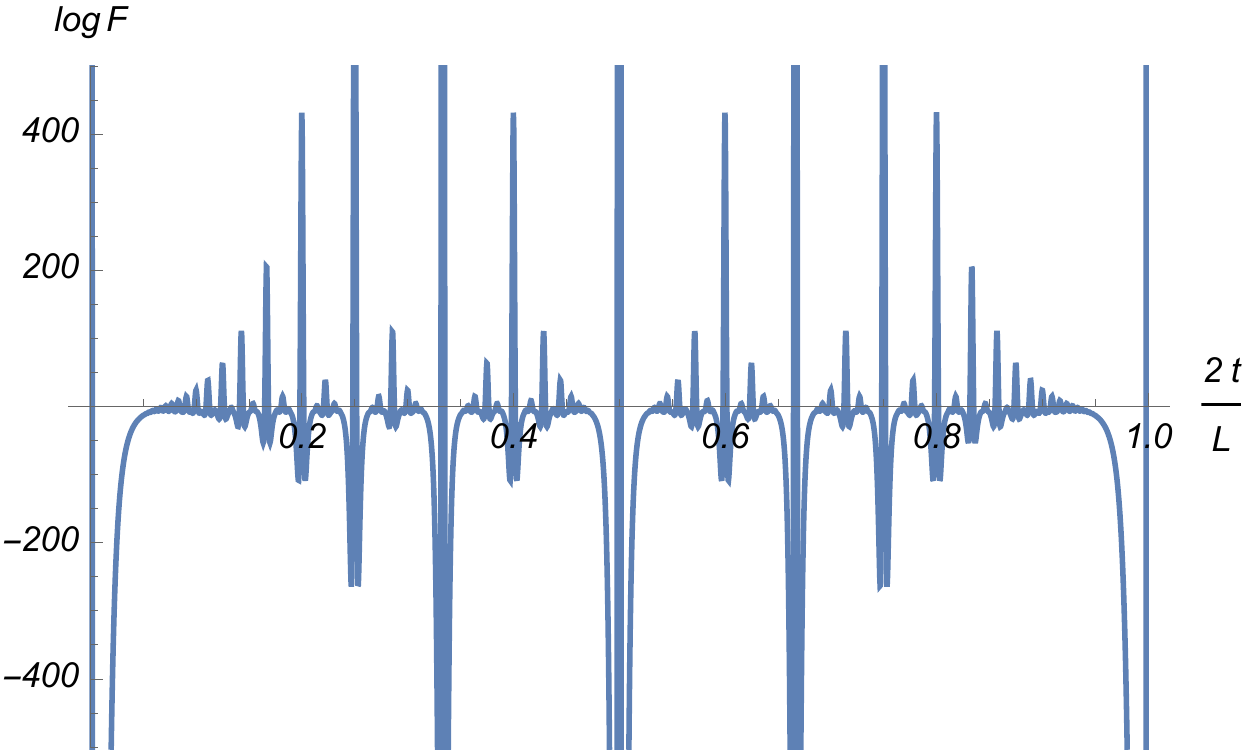}
\end{subfigure}%
\begin{subfigure}{0.5\textwidth}
\centering
\includegraphics[width=0.9\linewidth]{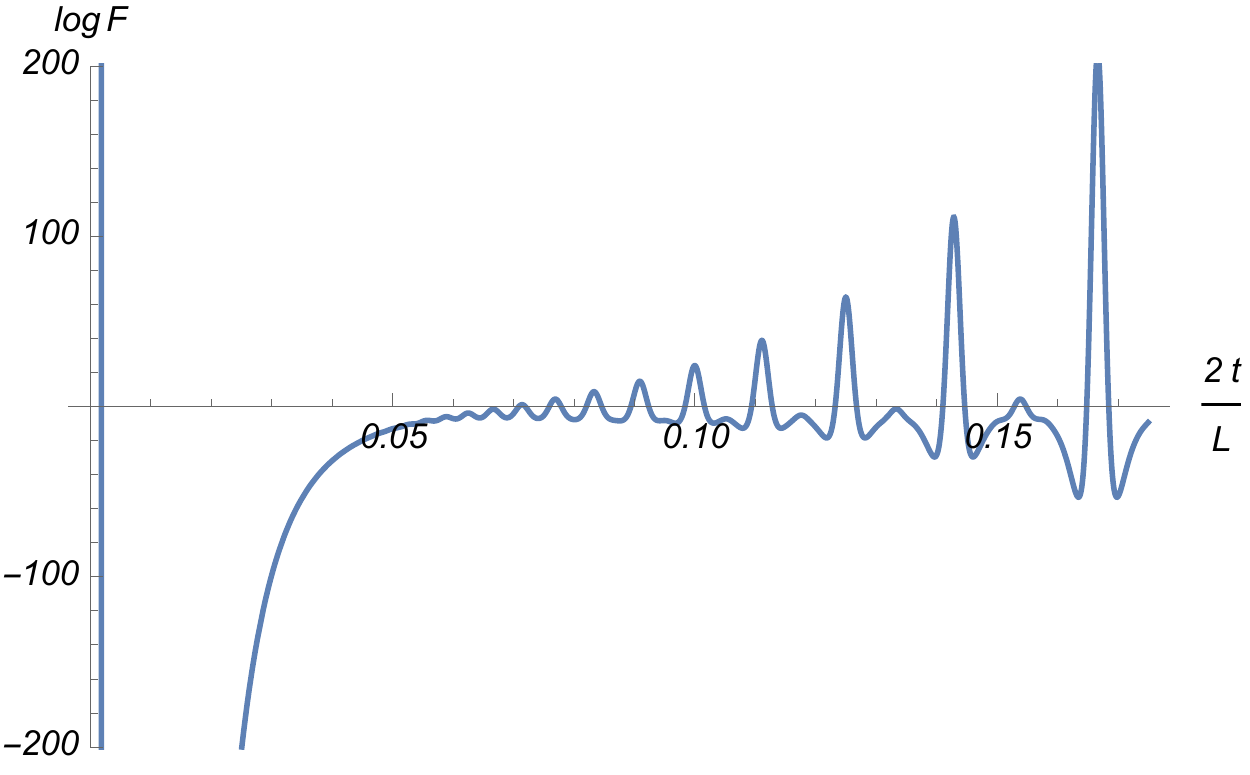}
\end{subfigure}
\caption{\small The same as Fig.~\ref{Fig4da} with $2\pi\beta/L$\ $=0.02$. Many more partial revivals are now visible.
Each is a scaled echo of the initial oscillation.
In the right hand figure the scale is expanded to show how the initial oscillatory decay morphs into partial revivals at $2t/L\approx1/m$.}\label{Fig4db}
\end{figure}

Having this result in hand we may now compute the return amplitude by setting $q=e^{2\pi(\beta+2it)/L}$.
We see immediately that there are complete revivals when $2t/L$ is an integer for $d$ even, and when $t/L$ is an integer for $d$ odd. Some plots of $\log{\cal F}(t)$ for $d=4$ are shown in Figs.~(\ref{Fig4da}, \ref{Fig4db}).

They look remarkably similar to the 2d case in Fig.~\ref{Fig2d}, except that we can clearly see the initial oscillation and also its echo near the full revival as predicted from the Casimir term. However these figures also show interesting structure near other rational values of $t/L$. This is to be expected given the form of the denominators in (\ref{Zd}), but the precise dependence near these points may, for even $d$, be decoded as a result of the properties of (\ref{Zd}) under the modular group.

\subsection{Modular properties and partial revivals.}
Let us define $\delta$ by $q^{-2\pi\delta}$. (In the literature of the modular group $\delta$ is usually written as $-i\tau$, but we have already used $\tau$ as imaginary time.) The modular group is generated by $T:\delta\to \delta-i$, which corresponds to the fundamental revival period, and $S: \delta\to 1/\delta$. It is the behavior of $Z^c_d$ under this element which is by no means obvious.  For simplicity we consider $d=4$ and refer the reader to \cite{JC91} for all details. 

There it was shown that if we define 
\be\label{Jd}
J_4(\delta)\equiv-\delta^2\frac\partial{\partial\delta}\log Z_4(e^{-2\pi\delta})-6(2\pi)^{-3}\zeta(4)\delta^{-2}\,,
\ee
then
\be\label{mod4}
J_4(\delta)=J_4(1/\delta)\,,
\ee
that is, $J_4$ is invariant under $S$.  The reason for this may be seen formally from (\ref{Zlog}):
$$
\log Z_4(\delta)\propto\sum_l\sum_n(l+1)^2\log\big((l+1)^2+n^2/\delta^2\big)\,,
$$
so
$$
-\delta^2(\partial/\partial\delta)\log Z_4(\delta)\propto\sum_l\sum_n\frac{(l+1)^2n^2}{(l+1)^2\delta+n^2/\delta}\,,
$$
so that the symmetry $\delta\to1/\delta$ corresponds to the formal interchange $l+1\leftrightarrow n$. The regularization then leads to the additional term in (\ref{Jd}).

Since $Z_4(1/\delta)=1+O(e^{-2\pi/\delta})$ 
as $\delta\to0$, we see that
$$
-\delta^2\frac\partial{\partial\delta}\log Z_4(e^{-2\pi\delta})\sim 6(2\pi)^{-3}\zeta(4)\big(\delta^{-2}-\delta^2\big)
+O(e^{-2\pi/\delta})\,,
$$
so
$$
\log Z_4(e^{-2\pi\delta})\sim 6(2\pi)^{-3}\zeta(4)\big(\delta^{-3}+3\delta\big)+O(e^{-2\pi/\delta})\,.
$$
 This not only gives the correct Casimir term in the limit $\delta=\beta/L\ll1$ (with a coefficient $\sigma=6(2\pi)^{-3}\zeta(4)$ equal, up to trivial factors, to the usual Stefan-Boltzmann constant), but also shows that if we include the ground state energy $E_0=-3\sigma(2\pi/L)$ in the definition of $Z_4$, its modular properties 
under $S$ are simpler, just as for $d=2$.
We also see that the corrections are then $O(e^{-2\pi/\delta})$, as argued earlier.
 
 Recalling that in the quench problem $\delta=(\beta+2it)/L$, we now use the modular properties to investigate the behavior near rational values of $2t/L$. For simplicity consider  $2t/L\approx1/m$. Setting $\epsilon=2t/L-1/m$, we have $\delta=(\beta/L)+i/m+i\epsilon$, so
 $$
 \delta^{-1}=\big((\beta/L)+i/m+i\epsilon\big)^{-1}\approx -im+(\beta/L)m^2+i\epsilon m^2\,.
 $$
 Thus the modular symmetry (\ref{Jd},\ref{mod4}) implies, to lowest order in $\beta/L$ and $\epsilon$
 $$
 -(-im)^{-2}(\partial/\partial\delta)\log Z_4(e^{-2\pi\delta})\sim(\partial/\partial\delta)\log Z_4(e^{-2\pi/\delta})\,.
$$
 Using now the symmetry of the last term under under $\delta^{-1}\to\delta^{-1}+im$, the last expression is given by the Casimir limit, that is
$\log Z_4(e^{-2\pi(\delta+im)})\propto(\delta+im)^{-3}$. Finally this implies for the logarithmic return amplitude
$$
{\cal F}(2t/L\approx1/m+\epsilon)\sim m^{2}{\rm Re}[((\beta/L)m^2+i\epsilon m^2)^{-3}]\,,
$$
so that the behavior near this partial revival is an inverted echo of the initial decay, suppressed by a factor
$m^{-4}$. Note that this is a stronger attenuation than in 2d, where it is only $m^{-2}$ \cite{C1}.
This will be observable only if $(\beta/L)m^2\ll1$, that is, for a fixed $\beta/L$, only revivals with denominator $\lesssim (L/\beta)^{1/2}$ should be seen. 

These results may be generalized to arbitrary rational values of $2t/L$, using its continued fraction representation which corresponds to successive applications of the elements $S$ and $T^m$ of the modular group. Each action of $S$ suppresses the signal by a factor similar to the above. 

All the above features are illustrated in the plots in Fig.~\ref{Fig4db}.

\begin{figure}
 \centering
\includegraphics[width=0.6\textwidth]{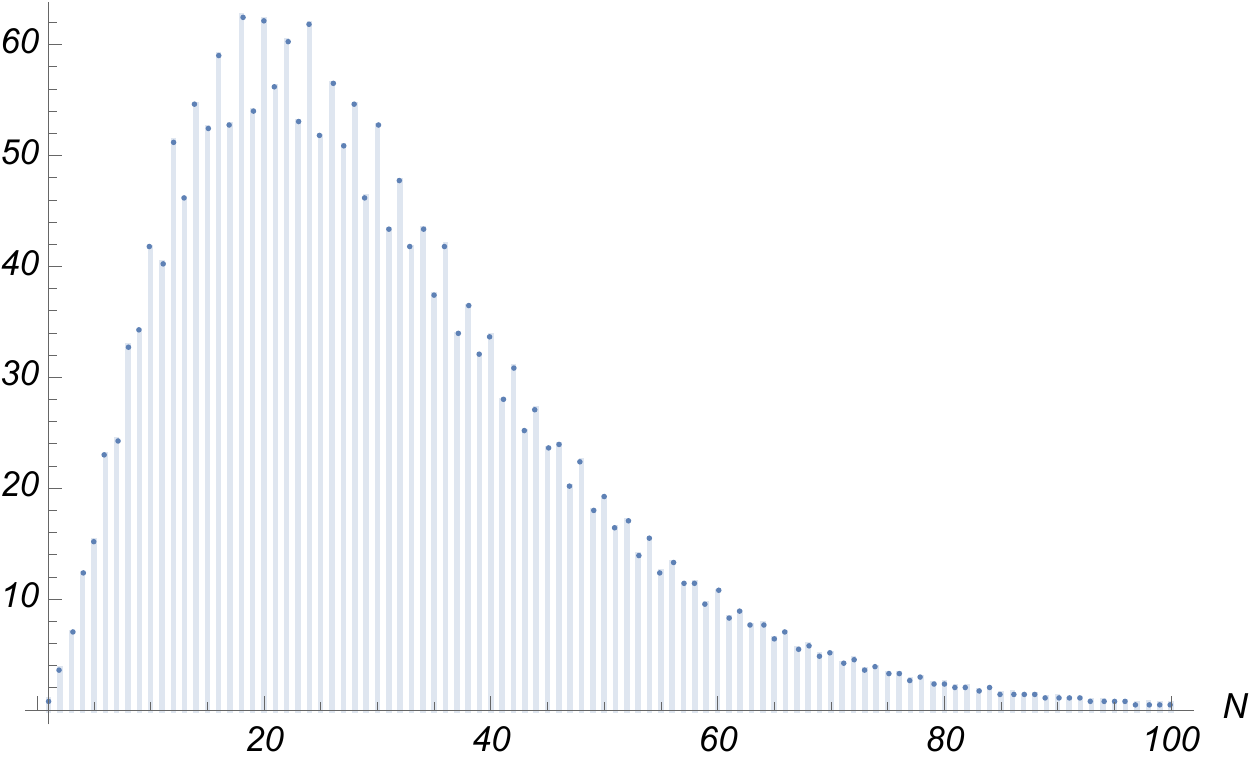}
\caption{\small The power spectrum of Fig.~\ref{Fig4db}. $N$ labels multiples of the fundamental frequency $4\pi/L$.
The apparent convergence to an exponential is an illusion since $\sigma_3(N)$ behaves erratically on larger scales.}\label{FigPS}
 \end{figure}

Given the complicated behavior seen in $\log{\cal F}(t)$ it is interesting to consider its power spectrum (Fourier series). Indeed we have, for the numerator in (\ref{Fratio}),
$$
|\log Z_4(\delta=(\beta+2it)/L)|={\rm Re}\,\sum_{l=0}^\infty(l+1)^2\sum_{j=1}^\infty j^{-1}e^{-2\pi j(l+1)(\beta+2it)/L}\,,
$$
so that the coefficient of $\cos(4\pi Nt/L)$ is
$$
e^{-2\pi N\beta/L}\sum_{(l+1)j=N}\frac{(l+1)^2}{j}=N^{-1}\sigma_3(N)e^{-2\pi N\beta/L}\,,
$$
where $\sigma_k(N)=\sum_{n|N}n^k$ is the sum of the $k$th power of the divisors of $N$. An example of this power spectrum is shown in Fig.~\ref{FigPS}.

 \begin{figure}[ht]
\centering
\begin{subfigure}{0.5\textwidth}
\centering
\includegraphics[width=0.9\linewidth]{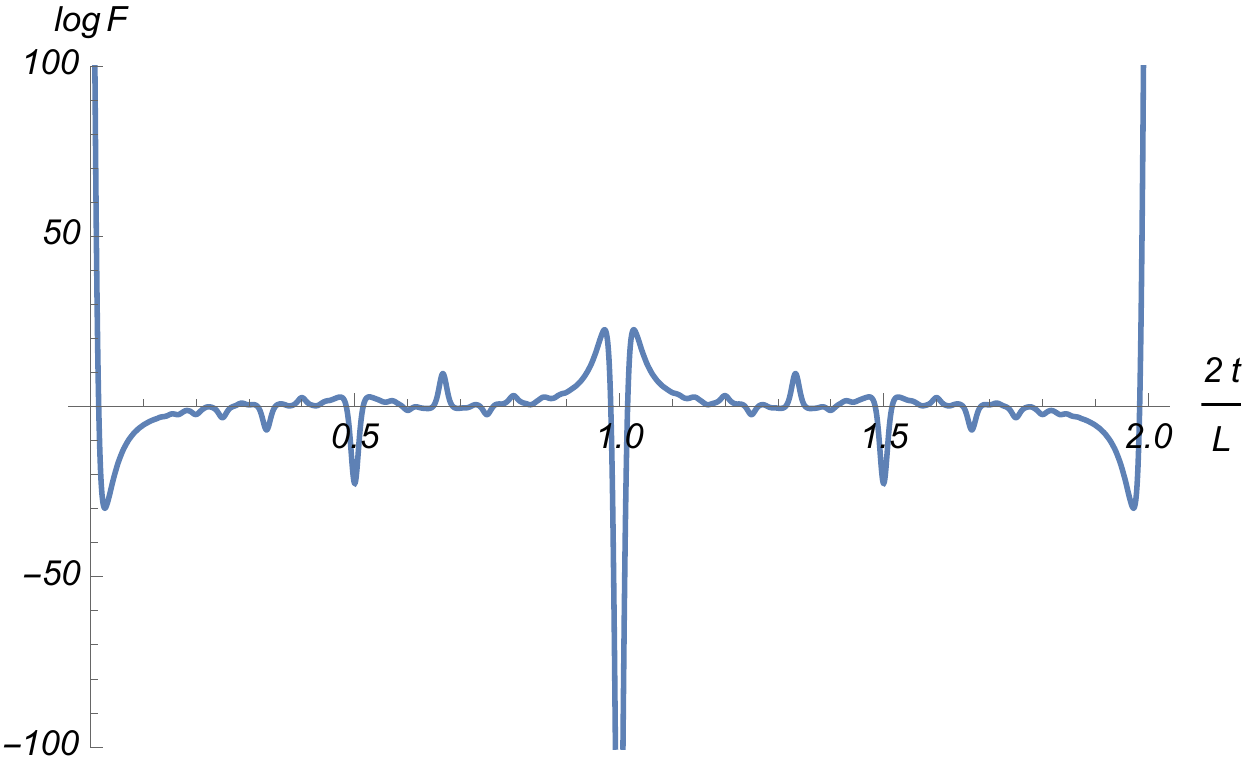}
\caption{\small $2\pi\beta/L=0.1$}
\end{subfigure}%
\begin{subfigure}{0.5\textwidth}
\centering
\includegraphics[width=0.9\linewidth]{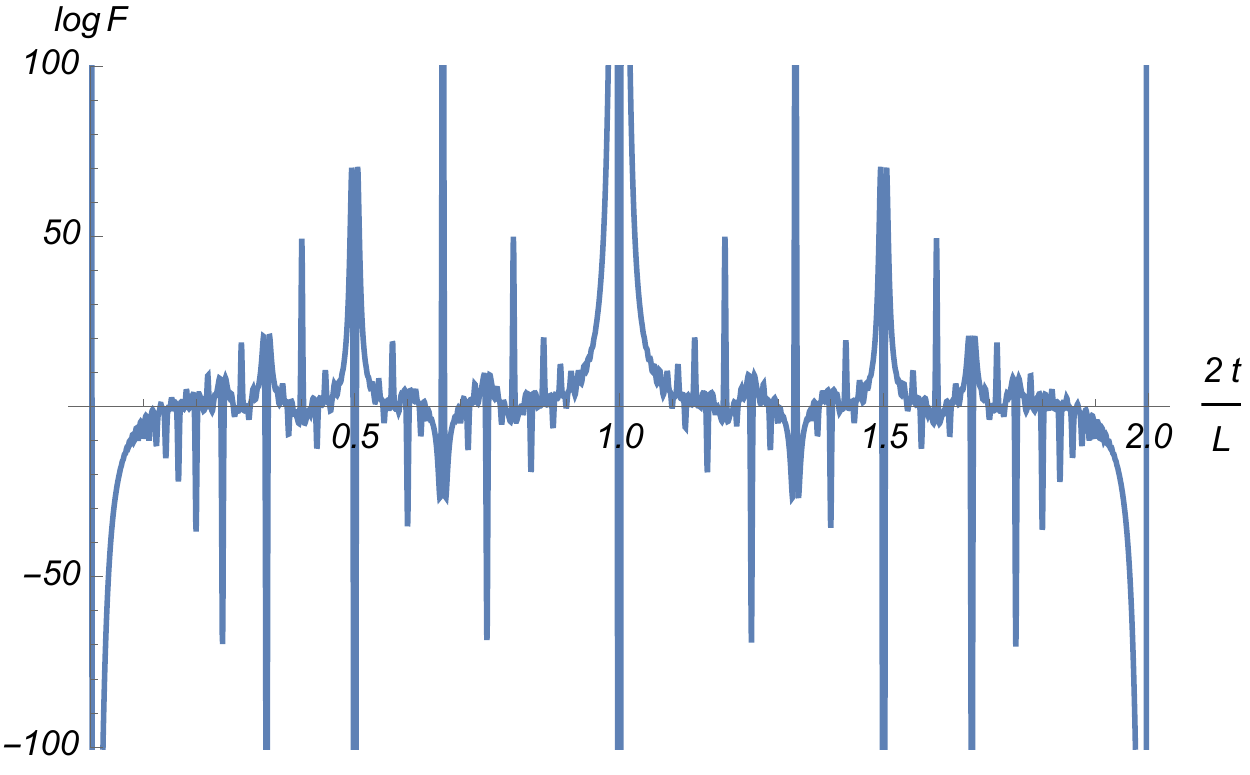}
\caption{\small $2\pi\beta/L=.02$}
\end{subfigure}
\caption{\small Logarithmic return amplitude for a conformally coupled free scalar field on $S_2$. Full revivals now occur only at even integer values of $2t/L$, and at odd integers there is destructive interference between even and odd values of $l$. Note that the peak and its accompanying oscillation is inverted there, and also at the echoes at some of the partial revivals.} \label{Fig3d}
\end{figure}
 
 We now turn to the case of odd $d$, using $d=3$ as an example. It may be seen immediately that full revivals now occur only when $2t/L$ is an even integer, and at odd integers there should in fact be a minimum due to destructive interference between odd and even values of $l$. This is seen in Fig.~\ref{Fig3d}. This is typical of the discrete but irregular power spectrum of an integrable system.
 
 The feature that the oscillation at $2t/L=1$ is inverted relative to that at the full revivals may be seen by studying the partition function near this value. Indeed, setting $\delta\to\delta-1$, that is $q\to e^{2\pi i}q$, we have
 $$
 Z_3( e^{2\pi i}q)=\prod_l\frac1{\big(1+q^{l+1/2}\big)^{2l+1}}\,.
 $$
 This is the inverse of the partition function for a free fermion on $S_2$. Thus the peak in $\log Z$ is inverted, and its overall size it determined by the Casimir energy for a fermion rather than a boson, which is smaller. This may be seen in Fig.~\ref{Fig3d}.

\begin{figure}[ht]
 \centering
\begin{subfigure}{0.5\textwidth}
\centering
\includegraphics[width=0.9\linewidth]{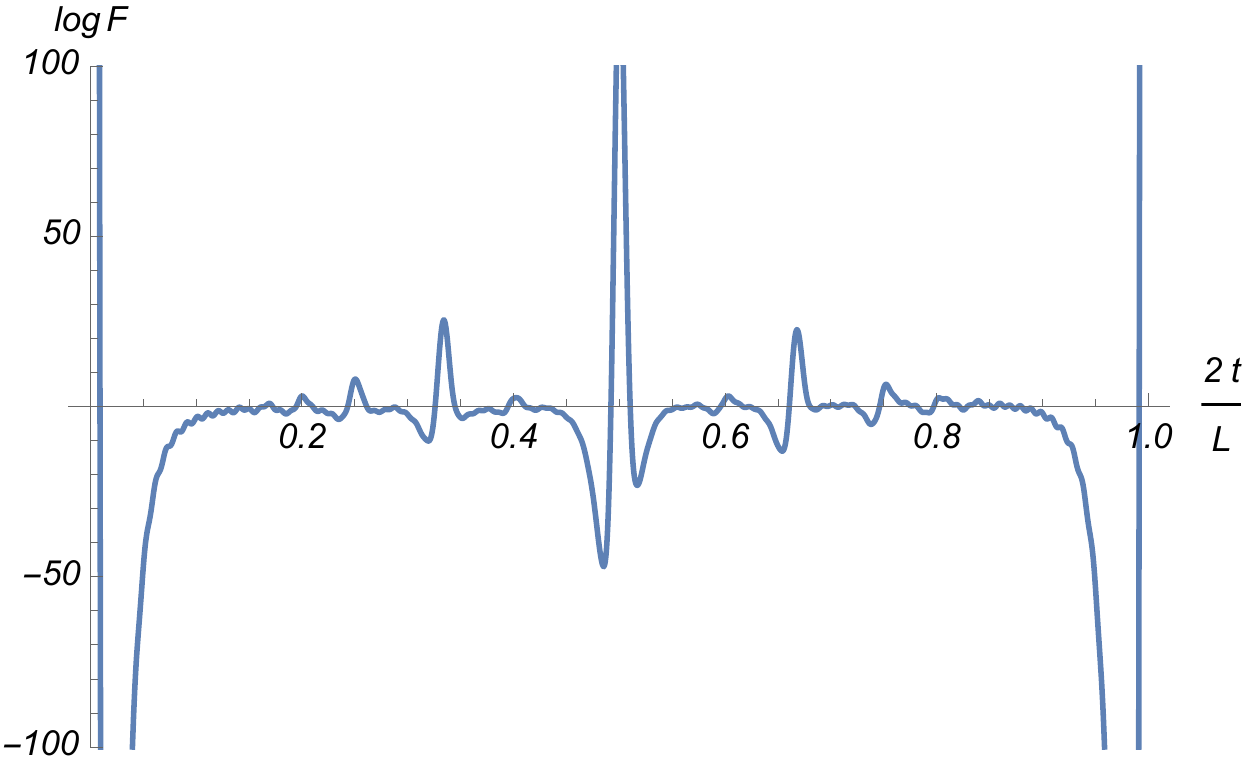}
\caption{\small $2t/L\leq1$}
\end{subfigure}%
\begin{subfigure}{0.5\textwidth}
\centering
\includegraphics[width=0.9\linewidth]{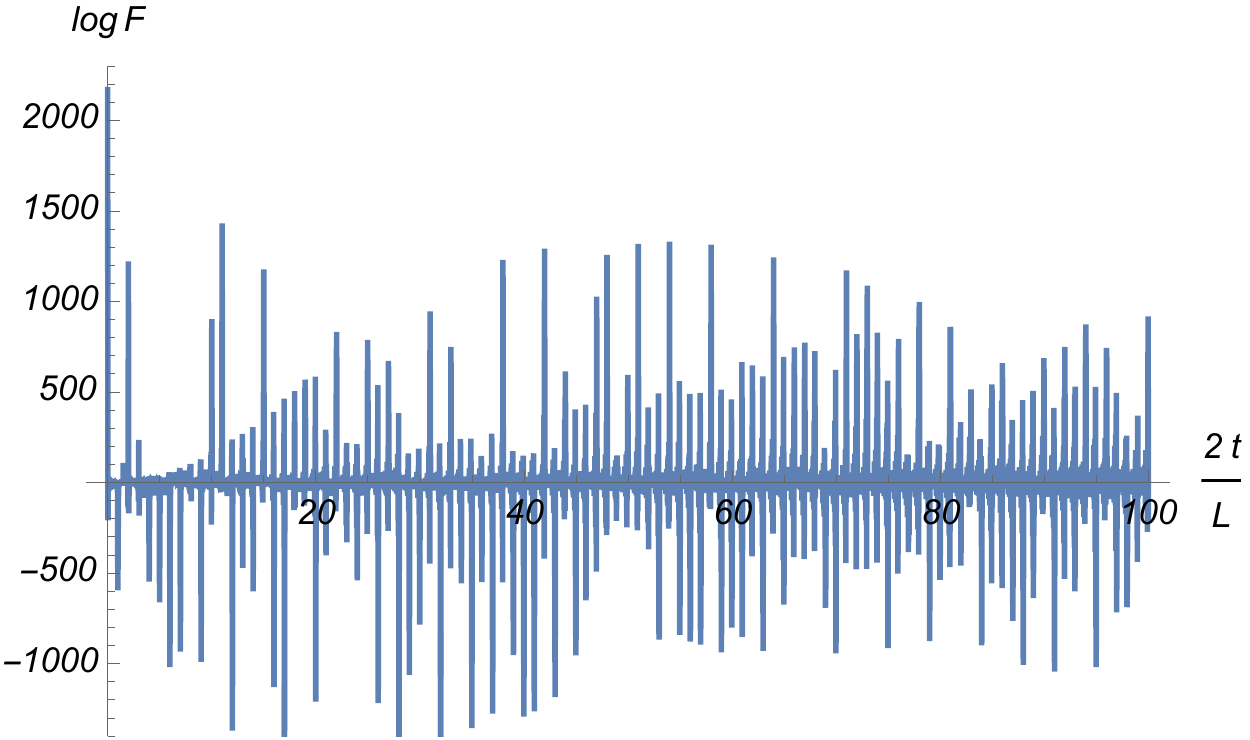}
\caption{\small $2t/L\leq 10^2$}
\end{subfigure}
\caption{\small Non-conformally coupled case on $S_3$ with $\lambda=0.5$ and $2\pi\beta/L=0.1$. There are still partial revivals for $2t/L<1$ although they are asymmetric and irregular. The RH figure shows that there are still strong  although incomplete revivals at larger values of $2t/L$.} \label{FigNCa}
\end{figure}

\begin{figure}\label{FigNCb}
\centering
\includegraphics[width=0.6\textwidth]{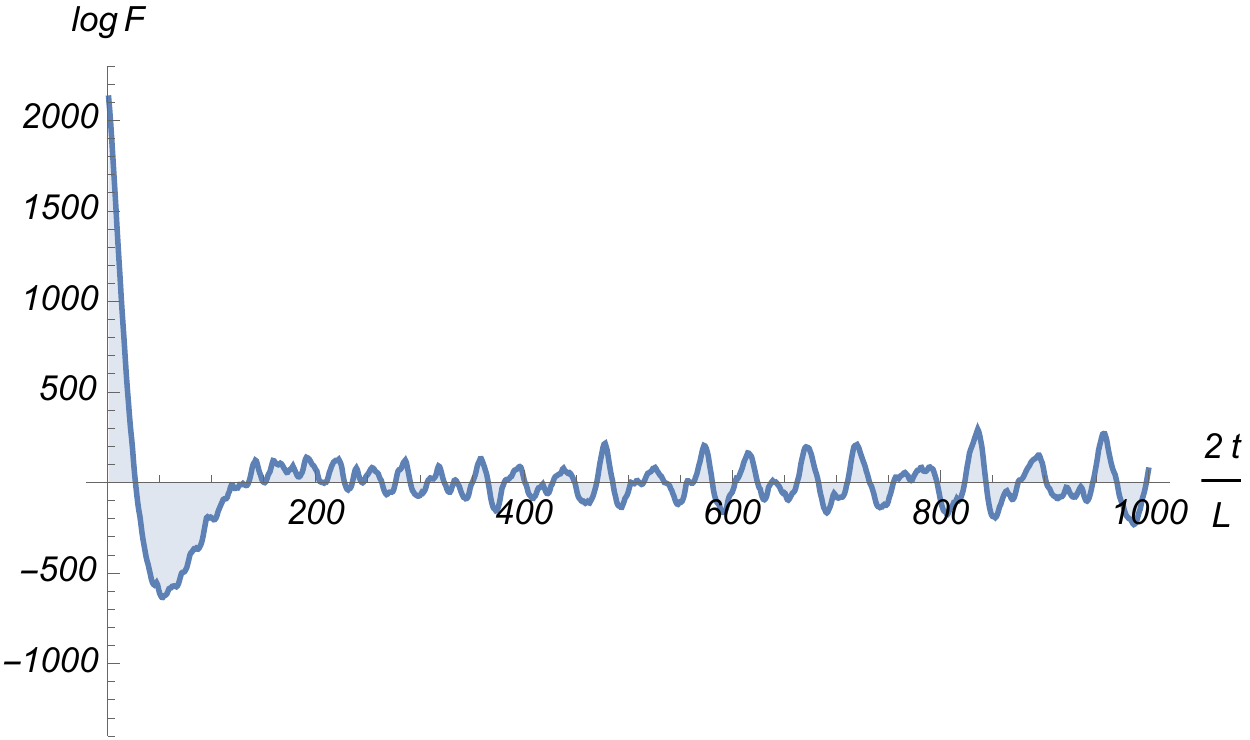}
\caption{\small As in Fig.~\ref{FigNCa}(b) with values only at integer $2t/L\leq 10^3$ shown.}\label{integervalues}
\end{figure}

 Finally we consider the case of a non-conformally coupled theory.  Taking $d=4$ as an example, this corresponds to a shift in the single-particle energies
  $l+1\to\sqrt{l(l+2)+\lambda}$ with
 $\lambda\not=1$. Because of the factor $e^{\sqrt{l(l+2)+\lambda}(4\pi it/L)}$ there will no longer be exact revivals at integer $2t/L$.   This may be seen in Fig.~\ref{FigNCa}. In fact the strong partial revivals do not occur at integer values of $2t/L$ (as may be seen in Fig.~\ref{integervalues}) but appear to be quite chaotic. 
 On the other hand, the visible partial revivals for $2t/L<1$ are scarcely affected apart from being asymmetrical. 
 
\section{Discussion}

We have investigated the behavior of the return amplitude ${\cal F}(t)=|\langle\Psi(t)|\Psi(0)\rangle|$ in CFTs on a compact spatial manifold $M$, in space-time dimensions $d>2$, following a quench from a state of the particular form (\ref{ccstate}). For a general CFT, the initial fall-off of ${\cal F}(t)$ is universal with an amplitude determined by the coefficient of the Casimir energy. For $d=3$ and $4$ it exhibits a single oscillation before decaying to a plateau value which is exponentially small in the volume of $M$. For a general CFT it remains there only for times $t\lesssim O((\beta L)^{1/2})$, but for holographic CFTs this plateau should persist up to $t=O(\sigma L)$ where $\sigma\gg1$ is again the Casimir coefficient (proportional to the Stefan-Boltzmann constant of the CFT.) 

We then considered the case of a  conformally coupled free scalar CFT when $M=S_{d-1}$. In this case, as for rational 2d CFTs, there are complete revivals where ${\cal F}=1$ at integer $2t/L$. There are also partial revivals
at rational values of $2t/L$, which, for $d=4$, may be understood in terms of the modular properties of the partition function.

\begin{figure}[ht]
 \centering
\begin{subfigure}{0.5\textwidth}
\centering
\includegraphics[width=0.9\linewidth]{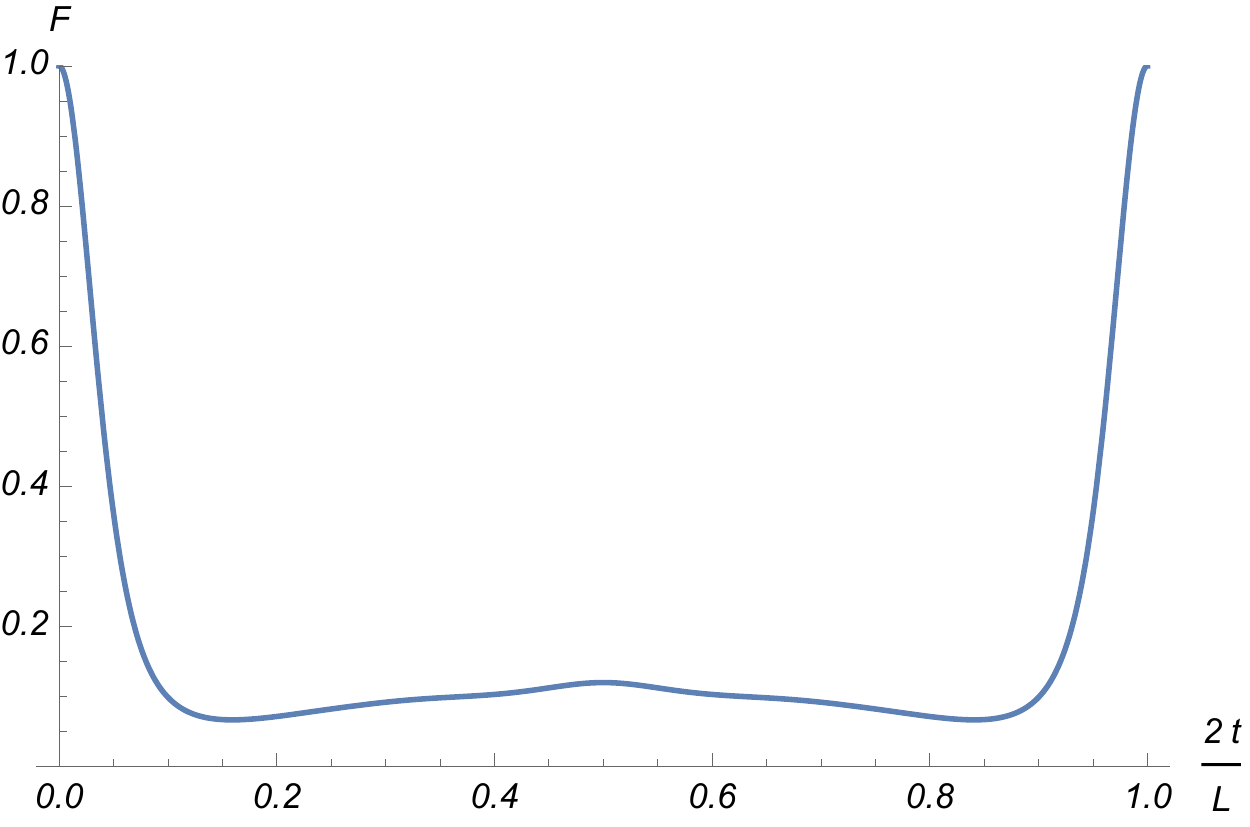}
\caption{}\label{Fig4dFa}
\end{subfigure}%
\begin{subfigure}{0.5\textwidth}
\centering
\includegraphics[width=0.9\linewidth]{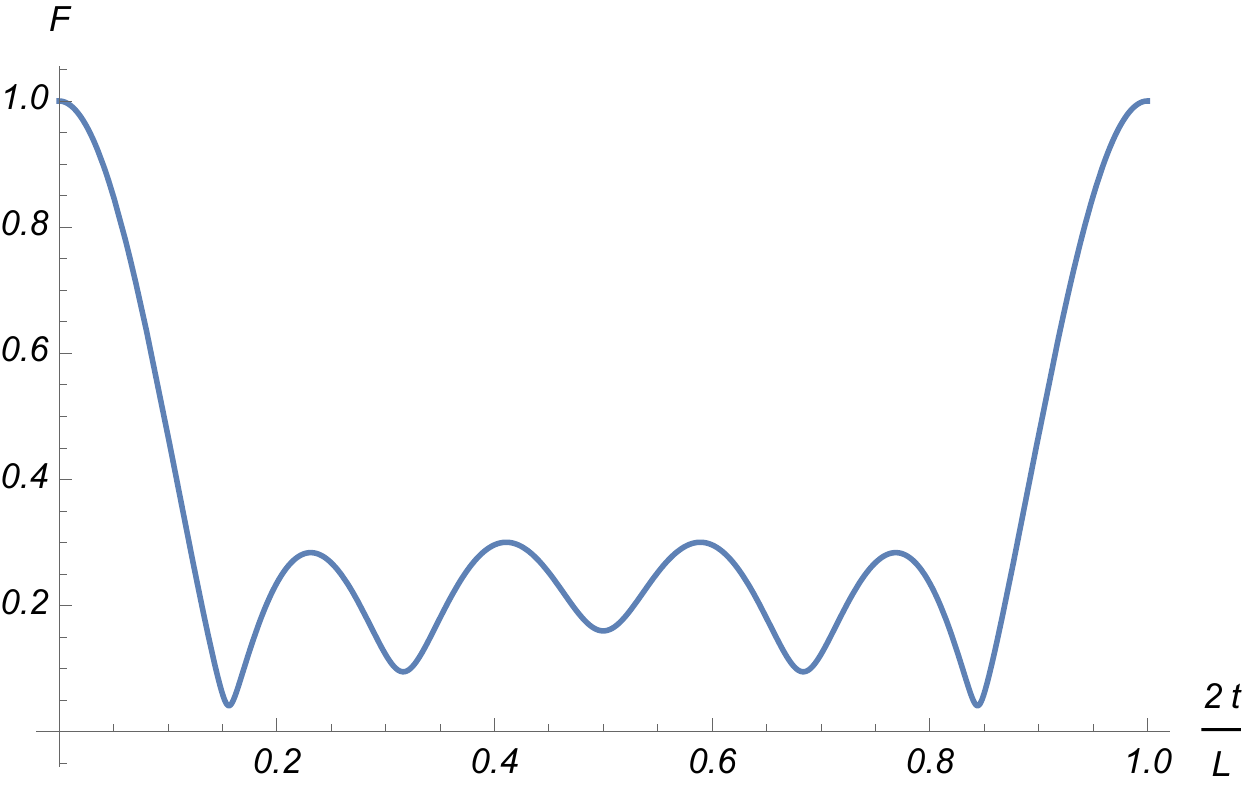}
\caption{}\label{Fig4dtrunc}
\end{subfigure}
\caption{\small(a) The normalized return probability  ${\cal F}(t)$ on $S_3$ when $2\pi\beta/L=1$. Although the complete revival at $2t/L=1$ is obvious, there are only slight indications of the partial revivals at $(\frac13,\frac12,\frac23)$.
(b) The same quantity with the series in (\ref{op}) truncated to $\Delta\leq4$.}
\end{figure}

 We should stress that the pervious illustrations are of the \em logarithmic \em return amplitude $\log{\cal F}$. Partial revivals are all exponentially suppressed in $L/\beta$, and to see any features in ${\cal F}$ itself it is necessary to consider smaller values of $L/\beta$, when the justification for considering the state (\ref{ccstate}) is less strong. In (\ref{Fig4dFa}) we show the actual return amplitude on $S_3$ when $2\pi\beta/L=1$. (It is important not to take this quantity too small, or the partial revivals are too small, not too large, when the ground state dominates the dynamics and there are no revivals either.)

 It is also interesting to consider $\beta/L=O(1)$, since the expansion of (\ref{op}) is then an expansion in powers of $e^{-2\pi}$ and it is instructive to see how much information can be gained by considering only the lowest few values of the scaling dimensions, which may be all that is approximately known for non-trivial CFTs. 
 In Fig.~(\ref{Fig4dtrunc}) we show the same plot with the series (\ref{op}) truncated to scaling dimensions $\Delta\leq4$. We see that the agreement with Fig.~(\ref{Fig4dFa}) is very poor, despite the fact that for this value of $\beta/L$ the series converges quickly when $t=0$, and the truncation actually overemphasizes the partial revivals. 

 It is also important to discuss how realistic the initial state (\ref{ccstate}) is. As discussed earlier and also in Ref.~\cite{JCGGE}, more general states may be considered by acting with other operators on the conformal state $|D\rangle$. In the case of a free field theory on $S_{d-1}$ these would be nonlinear in the number operators 
  $n_{lm}=a^{\dag}_{lm}a_{lm}$. In the partition function these will have the effect of coupling different angular momenta and in general be difficult to analyze. Similarly one may consider perturbations of the CFT hamiltonian itself, such as were considered in \cite{JCGGE}. The simplest, although not the most general, scenario is to take into account both of these effects and consider (\ref{ret}) with now
  $$
  H= H_{CFT}+g\int_M\Phi d^{d-1}x\,,
  $$
  where $\Phi$ is an irrelevant scalar operator of dimension $d+\mu$ with $\mu>0$. To first order in $g$, dimensional analysis implies that the energy of an eigenstate of $H_{CFT}$ of energy $E=2\pi\Delta/L$ 
 will be shifted by a relative amount $O(gE^{\mu})=O(g(\Delta/L)^{\mu})$. We may therefore replace 
   $$
   \Delta\to\Delta\big(1+\tilde gL^{-\mu}\Delta^\mu\big)\,,
   $$
   where $\tilde g\propto g$.
   
This assumes, of course, that all the matrix elements of the perturbation between degenerate states are equal.
This happens to be the case for the $T\overline T$ perturbation in 2d discussed in \cite{C1}, but more generally we should treat this as an oversimplified model. It has the effect in (\ref{op}) of replacing
 $q^\Delta\to q^{\Delta(1+\tilde gL^{-\mu}\Delta^\mu)}$. Setting $q^{\tilde gL^{-\mu}}=e^{-x}$
 we may write, via a steepest descent estimate
 $$
  q^{\tilde gL^{-\mu}\Delta^{1+\mu}}=e^{-x\Delta^{1+\mu}}\propto \int e^{-x^{-1/\mu}u^{1+1/\mu}+u\Delta} du\,,
  $$
 up to unimportant constants. The partition sum is then
 $$
 \int e^{-x^{-1/\mu}u^{1+1/\mu}}\,Z(qe^u)du\,,
 $$
 where $Z(q)$ is the unperturbed sum (\ref{op}). Now recall that in the numerator of (\ref{Fratio}) 
 $q\to e^{-2\pi(\beta+2it)/L}$ so $x\to\tilde g L^{-\mu}(2\pi(\beta+2it)/L)$. This means that near a revival at $2t/L=n$, where $n$ is an integer $\geq1$, $(\beta+2it)/L$ gets smeared by an amount $\sim u$, where $u\sim x^{1/(\mu+1)}\sim   [\tilde g L^{-\mu}(\beta/L+in)]^{1/(\mu+1)}$. This has the effect of a broadening of the $n$th revival of order $n^{1/(\mu+1)}$, although the actual behavior is more complicated due to the phase factors. 
 
 Thus we see that in this simple model the effect of a small irrelevant perturbation is to broaden and dampen each successive revival. Of course this should be valid only as long as $\tilde gL^{-\mu}n\ll1$, after which higher order effects should eventually destroy any revivals and lead to conventional thermalization.
 
 \em Acknowledgement. \em The author thanks J.S.~Dowker for pointing out a minor error in the first version on this paper.

\section*{Appendix}
Here we show that in a free field theory the state $e^{-(\beta/4)H}|D\rangle$ thermalizes to a simple Gibbs distribution.
For simplicity we take $M={\mathbb R}^{d-1}$, since we expect to find thermalization only on scales $\ll L$. Consider the 2-point correlation function in the Heisenberg picture\\
 $$
 G(x',t';x'',t'')=\langle\Psi(0)|\phi(x',t')\phi(x'',t'')|\Psi(0)\rangle\,.
 $$
 This may be found by computing the 2-point function in the euclidean slab geometry $\tau\in[-\beta/4,\beta/4]$
 $$
 \langle\phi(x',\tau')\phi(x'',\tau'')\rangle_{\rm slab}\,,
 $$
 and continuing $\tau'\to it'$, $\tau''\to it''$. This may be computed by a number of methods: the simplest is the method of images whereby
 $$
 \langle\phi(x',\tau')\phi(x'',\tau'')\rangle_{\rm slab}=\sum_n\left[G_0(x'-x'',\tau'-\tau''+n\beta)-
 G_0(x'-x'',\tau'-\tau''+(n-1/2)\beta)\right]\,,
 $$
 where $G_0(x,\tau)\propto(x^2+\tau^2)^{-(d-2)/2}$ is the Green function in $R^d$. For even $d$ this sum may be performed explicitly using the analyticity properties in $\tau$. Here we record only the result for $d=4$ and $t'=t''$:
 $$
 G(x',t;x'',t) =\frac\pi{2\beta r}\left(2\coth(\pi r/\beta)-\tanh(\pi(r-2t)/\beta)-\tanh(\pi(r+2t)/\beta)\right)\,,
 $$
 where $r=|x'-x''|$.

The first term is simply the equal-time thermal Green function $G_\beta(x'-x'')$. In fact we see that, except in a narrow region where $|r-2t|=O(\beta)$, to a very good approximation, which improves as $r,t\to\infty$
$$
G(x',t;x'',t) \approx G_\beta(x'-x'')\,\Theta(2t-|x'-x''|)\,.
$$
This is the higher dimensional version of the light-cone, or horizon, effect first noted in \cite{CC}: once $2t-|x'-x''|\gg\beta$ the 2-point correlations are identical to those at finite temperature. This is also the case for the equal-time correlations of $\dot\phi$. Since, for a gaussian state, these uniquely determine the reduced density matrix of the interval $[x',x'']$ we conclude that this is equal to the reduced density matrix of a Gibbs ensemble.

\end{document}